# Analysis of QoS of VoIP Traffic through WiFi-UMTS Networks

Mahdi H. Miraz, Suhail A. Molvi, Maaruf Ali, Muzafar A. Ganie and AbdelRahman H. Hussein

*Abstract*—Simulation of VoIP (Voice over Internet Protocol) traffic through UMTS (Universal Mobile Telecommunication System) and WiFi (IEEE 802.11x) alone and together are analysed for Quality of Service (QoS) performance. The average jitter of VoIP transiting the WiFi-UMTS network has been found to be lower than that of either solely through the WiFi and the UMTS networks. It is normally expected to be higher than traversing through the WiFi network only. Both the MOS (Mean Opinion Score) and the packet end-to-end delay were also found to be much lower than expected through the heterogeneous WiFi-UMTS network.

## I. INTRODUCTION

DUE to the continued prevalence of mobile internet usage, it is becoming increasingly difficult to ignore the importance of using VoIP to make calls. With the continued proliferation of mobile and fixed telecommunication networks, such as UMTS, LTE and WiMAX, the digitized data has to traverse multiple networks. The user is now experiencing various types of signal degradations, such as latency and jitter whose values differ from that experienced over the PSTN (Public Switched Telephone Network). The main aim of the research is to find out to what extent the QoS of the VoIP traffic varies while traveling through the latest heterogeneous generation of networks. To achieve this aim, our objectives include: 1. Designing different network scenarios using the OPNET modeler; 2. To simulate them to assess their performance and 3. To analyze the results obtained through the simulation. The first two scenarios will consist of several VoIP clients exchanging data through UMTS-to-UMTS and WiFi-to-WiFi. The QoS components of VoIP traffic such as the: MOS, Availability, Throughput, Distortion, Cross-talk, , Attenuation, Link Utilization, Loss, Jitter, Packet end-to-end Delay, Packet Delay Variation and Echo - will be examined and analysed. In the third scenario, the VoIP traffic will travel through heterogeneous networks i.e. UMTS-to-WiFi. The results previously obtained from the homogeneous scenarios will then be analysed and compared with.

The paper has been organized according to the following sections: Section-1 is the introduction to the research. Section-2 consists of the background information and related technological terms. Section-3 comprises the "Literature Review". Section-4 contains the research methodology describing the simulation scenarios and the required configuration. Section-5 analyses the results which is then followed by the conclusion and guidelines for future work.

## II. BACKGROUND

### A. VoIP (Voice over Internet Protocol)

VoIP [1] is the transmission of packetized voice traffic over a digital data network. This mode of transmission being set up as a shared virtual circuit differs from the analogue non-shared reserved switched circuit connection. The packets now have the freedom to take any route and also arrive in any order. This efficient use of channels also means having to contend with disadvantages such as that the packet can experience delays or never arrive at all due to severe traffic congestion. The flexibility of multiple routing does give the advantage of cheaper or even free routing of VoIP traffic. Sending digital data as packets means that digitized multimedia information can now be transmitted over a common shared network.

Thus the mainly software based VoIP technology has the major advantage of inexpensive scalability compared to other telephony systems. However, since being a mainly software based system - it is vulnerable to increasing targeted attacks from hackers currently in the form of worms and viruses. Some security solutions to address this potential problem are discussed in [2].

3G has helped in the lead to convergence [3] especially between the internet, fixed and mobile services. Global mobile internet access anywhere, along with broadband multimedia usage, is becoming more prevalent, especially with the spread of WiFi, WIMAX and femtocells. This capability has been facilitated by UMTS Rel. 5 using the IP Multimedia Subsystem (IMS). The user data and signaling is transported using the IMS packet based control overlay network. The IETF (Internet Engineering Task Force) SIP

Manuscript received March 14, 2014; revised April 4, 2014. This work was supported in part by the University of Ha'il.

M. H. Miraz is with the Department of Computer Science and Software Engineering, College of Computer Science and Engineering, University of Ha'il, PO Box 2440, Ha'il, Saudi Arabia and also with Glyndŵr University, Wrexham, Wales, UK (e-mail: m.miraz@uoh.edu.sa and m.miraz@glyndwr.ac.uk).

S. A. Molvi is with the Department of Computer Science and Software Engineering, College of Computer Science and Engineering, University of Ha'il, PO Box 2440, Ha'il, Saudi Arabia (e-mail: s.molvi@uoh.edu.sa).

M. Ali is with the Department of Computer Science and Software Engineering, College of Computer Science and Engineering, University of Ha'il, PO Box 2440, Ha'il, Saudi Arabia (e-mail: maaruf@ieee.org).

M. A. Ganie is with the Department of Computer Science and Software Engineering, College of Computer Science and Engineering, University of Ha'il, PO Box 2440, Ha'il, Saudi Arabia (e-mail: m.ganie@uoh.edu.sa).

A. H. Hussein is with the Department of Computer Science and Software Engineering, College of Computer Science and Engineering, University of Ha'il, PO Box 2440, Ha'il, Saudi Arabia (e-mail: ar.hussein@uoh.edu.sa).

was adopted by 3GPP (Third Generation Partnership Project) for the call setup session of VoIP and other IP-based multimedia communication. The current UMTS and IEEE 802.11 WiFi networks fully support real-time services such as VoIP [4].

### B. SIP (Session Initiation Protocol)

The entire call setup procedure from the beginning to the end requires the exchange of the signaling and control information between the parties involved. This already complicated process becomes even more so where mobility is involved through a heterogeneous network. Different devices have different capabilities and the seamless flow of information between them needs *a priori* information about their capabilities before the full flow of information can take place. This is handled by the SIP [5] application layer control protocol working alongside the existing other protocols. The destination and source "user agents" discover each other and establish the seamless connection between them based on shared properties using SIP. SIP creates any necessary proxy servers where needed by intermediary nodes dealing with such events as: registration, invitation and other requests. Mobility features are also catered for including: name mapping; redirection and the maintenance of an external visible location invariant identifier [6]. SIP is both session and device independent handling a wide range of multimedia data exchange capabilities. SIP also has many more functionalities, such as: the ability to setup a multicast call; removal of participants; user availability, session setup, user capabilities, user location and session management.

### C. Quality of Service (QoS) Parameters

The flexibility of the data networks data handling capability over that of the traditional telephone services puts it at a great advantage [4]. The QoS for VoIP can be measured using several different types of subjective and objective measures, such as, the Mean Opinion Score (MOS), as shown in Table 1, the jitter and the end-to-end delay.

TABLE 1
THE SUBJECTIVE MEAN OPINION SCORE (MOS) [7].

| Quality Scale | Score | Listening Effort Scale |
|---|---|---|
| Excellent | 5 | No effort required |
| Good | 4 | No appreciable effort required |
| Fair | 3 | Moderate effort required |
| Poor | 2 | Considerable effort required |
| Bad | 1 | No meaning understood with effort |

The MOS is calculated using a non-linear mapped R factor [8] as shown in Eq. (1), below:

$$MOS = 1 + 0.035R + 7 \times 10^{-6} [R(R - 60)(100 - R)] \quad (1)$$

Where:
 $R = 100 - I_s - I_e - I_d + A$
 $I_s$: voice signal impairment effects;
 $I_e$: impairment losses suffered due to the network and codecs
 $I_d$: impairment delays particularly mouth-to-ear delay.

The '*jitter*' "is the variation in arrival time of consecutive packets" [10], evaluated over a period of time [7]. It is the signed maximum time difference in the one way delay over a set time period. Wherever buffers are used they can both over-fill and under-fill, resulting in packets being discarded. Let t(i) be the time transmitted at the transmitter and t'(i), the time received at the receiver, the jitter is then defined as:

$$\text{Jitter} = \max_{1 \leq i \leq n} \{[t'(n) - t'(n-1)] - [t(n) - t(n-1)]\} \quad (2)$$

The '*Packet end-to-end delay*' "is measured by calculating the delay from the speaker to the receiver [including the] compression and decompression delays" [9]. $D_{e2e}$, the total voice packet delay, is calculated thus:

$$D_{e2e} = D_n + D_e + D_d + D_c + D_{de} \quad (3)$$

where $D_n$, $D_e$, $D_d$, $D_c$ and $D_{de}$ represent the network, encoding, decoding, compression and decompression delay, respectively.

The International Telecommunication Union – Telecommunication (ITU-T) offers guidelines for these as shown in Table 2 [10].

TABLE 2
GUIDELINE FOR THE VOICE QUALITY [9].

| Network parameter | Good | Acceptable | Poor |
|---|---|---|---|
| Delay (ms) | 0-150 | 150–300 | > 300 |
| Jitter (ms) | 0-20 | 20–50 | > 50 |

The '*Packet Delay Variation*' (PDV) is an important factor in network performance degradation as it affects perceptual quality. Higher packet delay variation leads to congestion of the packets causing more network overheads. The PDV is the variance of the packet delay, given by:

$$PDV = \{ \sum_{i=1}^{n} ([t'(n) - t(n)] - \mu)^2 \}/n \quad (4)$$

Where; *μ* is the average delay of *n* selected packets.

### D. WiFi

The contention based IEEE 802.11x Wireless LAN or WLAN is derived from the non-wireless Ethernet 802.3 Local Area Network (LAN) access technology. Layer 1 (physical) and Layer 2 (data link) operate over two frequency bands of 2 GHz and 5 GHz. 802.11b (11 Mbit/s) and 802.11g (54 Mbit/s) are two common standards with a range of between 80-100 m. Contention will reduce the practical bitrates and affect the QoS for real-time traffic, especially VoIP. The large packet headers, for both the WiFi and VoIP, constrain the payload capacity and affect the QoS further in times of congestion. WiFi is cheap and comes as a standard feature on most network devices. WiFi can be found in: public transportation, public spaces, domestic and industrial applications [10].

### E. UMTS (Universal Mobile Telecommunication System)

The 3GPP (3rd Generation Partnership Project) UMTS takes its foundational architecture from the GSM network [11]. A list of these is as follows:

➢ The USIM or UMTS SIM card is backwards compatible in a GSM handset.
➢ The Signalling System Number 7 (SS7) MAP (Mobile Application Part) protocol of UMTS is an

- evolution of the GSM MAP.
- ➢ Similar but enhanced versions of the circuit and packet transmission protocols are utilised.
- ➢ Seamless mobility during hard handovers is facilitated by special procedures during circuit and packet switching.

3G (as shown in Fig. 1) offers many of these features as defined by IMT (International Mobile Telecommunication):

*Flexibility*

In terms of services, deployment, technology and air interfaces. The IMT-2000 standard allows for five radio interfaces based on three different access technologies: FDMA, TDMA and CDMA;

*Affordability*

To ensure global uptake by consumers, it must be cheap.

*Compatibility with existing systems*

3G must be able to work with previous generations to allow for global roaming as 2G systems still continue to grow and thrive globally.

*Modularity of Design*

Scalability allows for coverage of growing population centres including implementing new services with minimal cost.

*HSPA (High-Speed Packet Access)*

UMTS offers HSPA [12], which is a combination of HSDPA (High-Speed Downlink Packet Access) and HSUPA (High-Speed Uplink Packet Access). The 3GPP also defines, Enhanched HSPA or HSPA+ (3GPP Rel-7) and Rel-8 Long Term Evolution (LTE) – also known as Evolved UMTS Terrestrial Radio Access (EUTRA). 4G is currently being deployed [13].

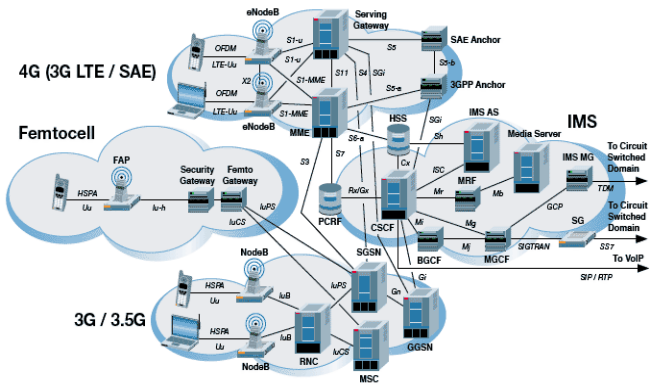

Figure 1: Sample 3G and 4G network.
(Source: http://www.ccpu.com/images/products/trillium/3g4gwireless1.gif)

III. LITERATURE REVIEW

A former study has shown that VoIP does adversely affect the throughput of both WiFi and WiMAX [10,14,15]. However, packet loss and jitter is only experienced in a WiFi network. Common OPNET network simulation software parameters used for performance studies include: "MOS, end-to-end delay, jitter, and packet delay variation" [7]. A study over High Speed Packet Access (HSPA) [14] has shown that not all VoIP implementations are the same as they have differing quality effects on the voice.

A strategy used to overcome the quality problems of congestion in WiFi when using VoIP is known as SQoSMA (Quality Assurance of Voice over WLANs) [16]. SQoSMA tries to overcome the bandwidth limitation problems by combining the control and data planes in order to mitigate congestion events by choosing the lowest bitrate adaptive audio codec. It then implements a call stopping procedure when necessary to reduce congestion even further.

A similar strategy [17] used an edge VoIP gateway between the Internet Cloud and the WLAN to select the most appropriate variable speech coding rate (16-64, Kbit/s) whilst maintaining overall QoS of the speech traffic.

Transmission delay reduction of VoIP traffic was implemented through a Transmission Control Protocol (TCP) - Friendly Rate Control (TFRC) algorithm based 802.11e network using EDCF (Enhanced Distributed Coordination Function)/HCF (Hybrid Coordination Function) scheme [17].

The use of routing and labelling may also be employed to help the rapid passage of the VoIP real-time traffic through the WLAN. The approach was to implement a distributive packet aggregation technique to reduce the time and number of hops through the network [18].

A study in subjective voice quality measure [19] looked at the E-Model and the PESQ (Perceptual Evaluation of Speech Quality) which combined the advantages of both to form the AdmPESQ (Advanced Model for Perceptual Evaluation of Speech Quality) measure. AdmPESQ is especially suitable for heterogeneous networks with differing packet losses and delay parameters.

VoIP continues to grow in popularity as it offers the services of the traditional Public Switched Telephone Network (PSTN) whilst incorporating added features at competitive rates. Many competing companies now offer this service over often heterogeneous networks, though this does affect the QoS. Furthermore VoIP is now the target of many computer hacker attacks, Materna [20] has categorized these into four:

1. Service Availability;
2. Service Integrity;
3. SPIT (Spam over Internet Telephony) and
4. Eavesdropping.

Network outage must be kept ideally at zero level as this may critically affect an organization, such as the vital operation of a hospital or nuclear power station. Thus "Service Availability Attacks" must be stopped. Downtime means financial loss and unplanned maintenance costs too. Thus the IP Telephony network must be robustly protected against all known forms of attacks such primarily the "Denial of Service" (DOS) types of attacks as well as from viruses and worms. Attacks will affect the service ranging for quality of service deterioration to total loss of service. Customers demand the highest level of service so voice quality has to be maintained.

VoIP services are more vulnerable than computers (protected more securely) to attacks due to their lower thresholds, immunity and higher sensitivity. Thus any attacks, for example a worm, will have a far greater impact on a VoIP network then on a traditional computer network. A computer may be slowed down but the VoIP network may totally crash.

The paper presents the results of VoIP traffic through a WiFi and a UMTS network and between them, especially in regards to the QoS.

## IV. RESEARCH METHOD

An academic environment often constrains research to simulations because of limited resources. However, much can be learned from this for future deployment into the real world for initial testing. Jack Burbank describes "Modeling and Simulation (M&S)" as an acute component in the "design, development and test and evaluation (T&E)" process. According to him, "It is almost always preferable to have insight into how a particular system, individual device, or algorithm will behave and perform in the real world prior to its actual development and deployment." [21] Simulation offers many advantages such as the opportunity to scale a network in a virtual environment thus saving considerable costs [22]. Comparison of technologies is easily achieved in simulations. Our project make use of the OPNET Modeler as it integrates a wide range of technologies and protocols [23], as well as comprises a "development environment" to facilitate M&S for various types of networks for our studies.

In the first scenario, two WiFi subnets, Hawaii and Florida, were deployed. These were configured with a SIP server credential connected through an IP cloud, shown in Fig. 2. In the second scenario, two UMTS subnets, New York and California, were used instead of the WiFi subnets. In the third scenario, one of the UMTS subnet (New York) was replaced by a WiFi subnet (Hawaii). The subnets implemented in the project are shown in Table 3.

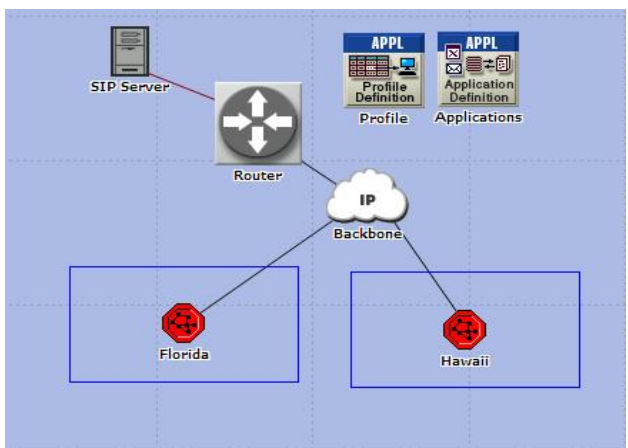

Fig. 2. WiFi Network Scenario

In the second scenario, two UMTS subnets, New York and California, were used instead of the WiFi subnets. In the third scenario, one of the UMTS subnet (New York) was replaced by a WiFi subnet (Hawaii). The subnets implemented in the project are as given in Table 3.

Workstation in both the UMTS and Wi-Fi network models are configured to run the VoIP application. This VoIP Application is defined to generate one voice frame per packet and to run as an 'Interactive Voice' service. This application is defined in the application profile to run in serial mode. Calls to workstations are based on random generation and are exponentially distributed with average duration of three (3) minutes. The call inter-arrival time are exponentially distributed. UMTS has two major divisions, namely the UMTS Terrestrial Radio Access Network (UTRAN) and the Core Network (CN), as shown in Fig 2. The UTRAN is a combination of two parts: the Radio Network Controller (RNC) and the Node-B.

TABLE-3
DEVICES OF THE SUBNETS DEPLOYED.

| Subnet Name | Scenario | Base Station Type | Work Station Type | Number of Work Station |
|---|---|---|---|---|
| Hawaii | WiFi | Wifi | Mobile | 4 |
| Florida | WiFi | Wifi | Mobile | 4 |
| New York | UMTS | UMTS | UMTS Workstation | 4 |
| California | UMTS | UMTS | UMTS Workstation | 4 |
| Hawaii | WiFi | WiFi | Mobile | 4 |
| California | UMTS | UMTS | UMTS Workstation | 4 |

The UTRAN handles all radio related functionalities. The CN is responsible for maintaining the subscriber data and for switching the voice and data connections.

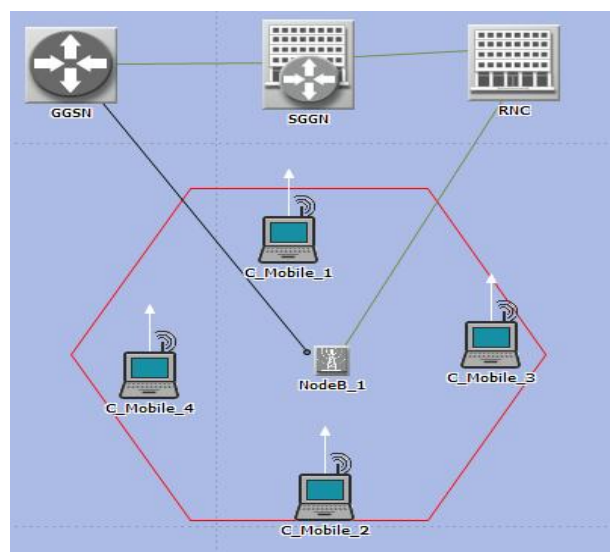

Fig. 3. Inside View of a UMTS Subnet.

## V. RESULTS AND DISCUSSION

The results were collected simulating all three scenarios for one (1) hour. Fig. 4a shows the average jitter of all three scenarios as overlaid graphs. Whereas Fig. 4b presents the average jitter graphs as stacked. The UMTS jitter has been found to be much higher (around 0.10 s) than that of the other two scenarios (around 0.0 seconds). In fact, the jitter graph for UMTS had no results for about the first five minutes. This is most probably due to the network convergence period. It is also to be noted that the mixed scenario presents negative jitter during this time due to the same reason. Negative jitter indicates that the packets arrived before the interval time during that period.

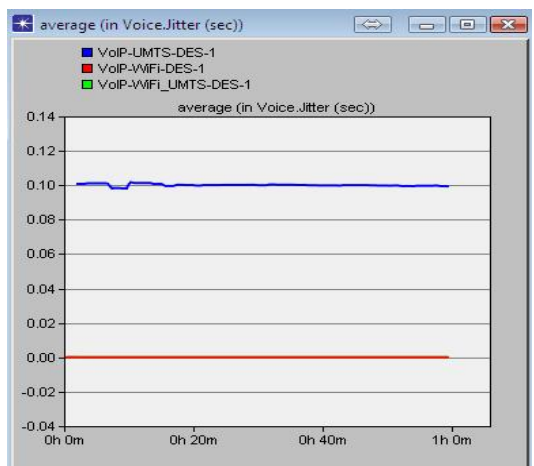

Fig. 4a. Average VoIP Jitter (Overlaid)

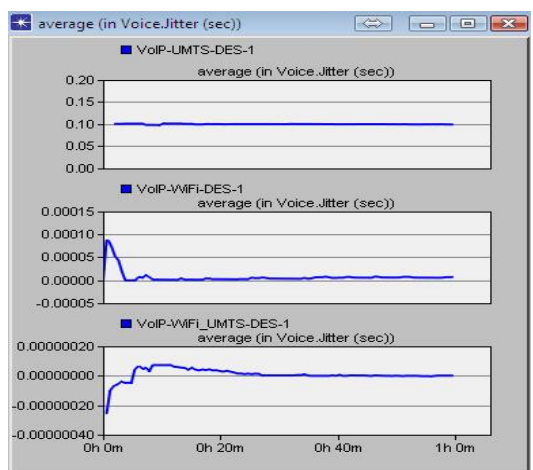

Fig. 4b. Average VoIP Jitter (Stacked)

It is quite interesting that the average jitter is not only very steady in both the WiFi and WiFi-UMTS scenarios but also remains nearly zero (0). As the simulation was configured to run on the basis of random call generation and there is no direct handover involved, the average jitter of the WiFi-UMTS scenario should preferably always stay somewhere in-between the WiFi and UMTS jitter times. This result, hence, is very thought provoking and demands further research.

This similar tendency has also been observed while comparing the MOS performance, as shown in Fig. 5. In terms of MOS, although the call generation was exponentially distributed, both the WiFi and WiFi-UMTS networks observe a similar level of performance. This remains very close to four (4), over the complete simulation period. UMTS, on the other hand, suffers from not only a lower level of MOS, but also an unsteady level. The MOS of these scenarios varied between: 1.5 to 2.8. So, considering MOS alone, it can be concluded that: 1) both the WiFi and the WiFi-UMTS networks surpass the UMTS network and 2) despite the MOS of the WiFi-UMTS network should ideally remain somewhere near the mid-point of the WiFi and UMTS MOS graphs, it exhibits a greater performance than that.

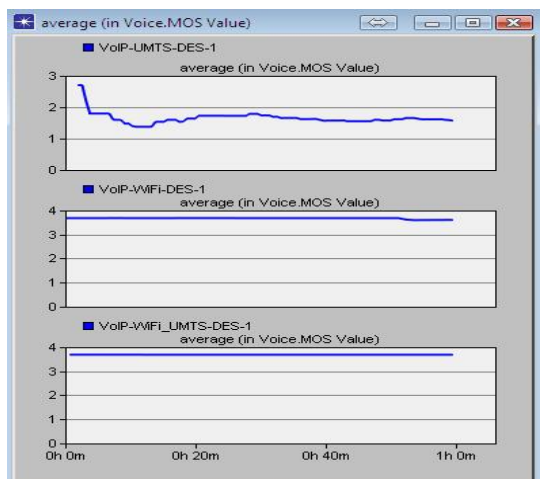

Fig. 5. Average MOS (Overlaid).

In terms of the packet end-to-end delay, unexpectedly, the WiFi-UMTS provides better services than either the WiFi or the UMTS networks, as shown in Fig. 4. The similar behaviour has also been observed while comparing the Packet Delay Variation, as shown in Fig. 6.

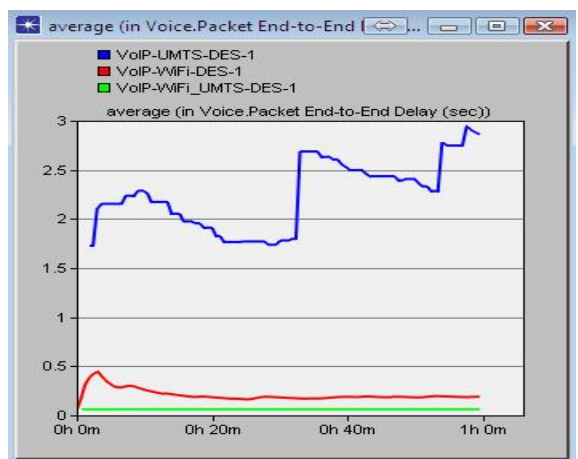

Fig. 7. Packet End-to-End Delay.

In terms of the packet end-to-end delay, surprisingly enough, the WiFi-UMTS provides better services than either the WiFi or the UMTS networks, as shown in Fig. 4. The similar behaviour has also been observed while comparing the Packet Delay Variation, as shown in Fig. 7.

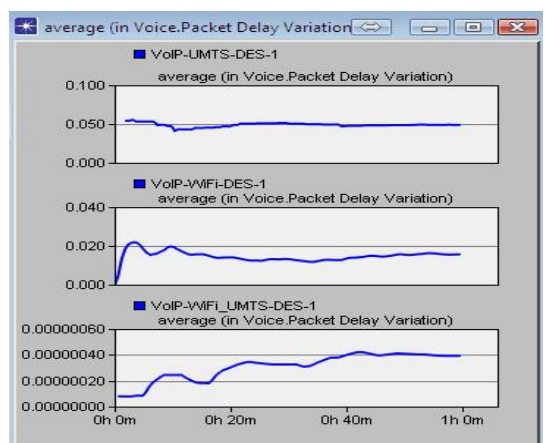

Fig. 6. Packet Delay Variation.

These results are very significant as it is expected that in the mixed (WiFi-UMTS) network case, the variable should remain in the middle under the ideal situation and might sometimes provide an inferior performance in the worst case scenario. Furthermore in depth research is required to identify the reasons behind this phenomenon.

## VI. CONCLUSION

The paper reported the preliminary findings of VoIP traffic through the WiFi, UMTS and WiFi-UMTS networks. To begin with, two different scenarios where deployed and tested where VoIP call generation and termination took place in the respective homogenous networks, these being the WiFi and UMTS. Later, an additional scenario was added to the project: VoIP calls being generated using the WiFi network terminated using the UMTS network and vice-versa.

While considering the Packet Delay Variation and the packet end-to-end delay, it has been found that the WiFi-UMTS heterogeneous network provides better services than the WiFi and UMTS networks separately. These results are unexpected as described in the discussion section of this paper. Further detailed research is definitely needed to be undertaken in order to identify the reasons behind this phenomena.

It is quite interesting that the average jitter is not only very steady in both the WiFi and WiFi-UMTS scenarios but also remains nearly zero (0). As the simulation was configured to run on the basis of random call generation with no direct handovers, the average jitter of the WiFi-UMTS scenario should preferably always stay somewhere in-between the WiFi and UMTS networks. This result, hence, is very perplexing and requires further investigation.

Considering the MOS, the UMTS network suffers not only from poor performance, but also remains unsteady. In addition to that the MOS performance of both the WiFi and WiFi-UMTS networks surpass that of the UMTS network and also the MOS of the WiFi-UMTS network exhibits a greater performance than expected.

For its future work plan, the project aims to include other networks technologies and techniques covering: CDMA, EDGE, GSM, GPRS, LTE and 4G. The effect of VoIP negotiating both a non-heterogeneous and heterogeneous network will be of one particular major focus of this ongoing research initiative. The study will be further extended to investigate the effect on other parameters such as on the: throughput, queuing delay and the packet drop rate. The network parameters will be carefully scrutinized for their optimization to improve the overall efficiency.